\documentclass[aps,prl,twocolumn,floatfix,superscriptaddress,longbibliography]{revtex4-1}

\usepackage[dvipsnames]{xcolor}
\usepackage{amsmath}
\usepackage{graphicx}
\usepackage{color}
\usepackage{textcomp}
\usepackage[utf8]{inputenc}
\newcommand{\one}{$\langle 111\rangle$}
\newcommand{\bto}{BaTiO$_3$}
\usepackage{hyperref}
\hypersetup{
    colorlinks=true,
    linkcolor=blue,
    filecolor=magenta,      
    urlcolor=cyan,
}

\usepackage{bibunits}
\defaultbibliography{BTO}
\defaultbibliographystyle{apsrev4-1}

\newif\ifgpint\gpinttrue 
\newif\ifmkint\mkinttrue 
\usepackage[normalem]{ulem} 

\begin{document}
\begin{bibunit}
\title{A microscopic picture of paraelectric perovskites from structural prototypes}
\author{Michele Kotiuga}
\email{michele.kotiuga@epfl.ch}
\affiliation{Theory and Simulation of
  Materials (THEOS) and National Centre for Computational Design and
  Discovery of Novel Materials (MARVEL), \'Ecole Polytechnique
  F\'ed\'erale de Lausanne, CH-1015 Lausanne, Switzerland}

\author{Samed Halilov}
\affiliation{Designed Material Technologies, LLC, P.O. Box 14548, Richmond, VA 23221-9998, USA}

\author{Boris Kozinsky}
\affiliation{John A. Paulson School of Engineering and Applied Sciences, Harvard University, 29 Oxford Street, Cambridge, MA 02138, USA}  
\affiliation{Robert Bosch LLC, Research and Technology Center, Cambridge, Massachusetts 02139, USA}  
  
\author{Marco Fornari}
\affiliation{Department of Physics and Science of Advanced Materials Program, Central Michigan University, Mt. Pleasant, Michigan 48859, USA}

\author{Nicola Marzari}
\affiliation{Theory and Simulation of
  Materials (THEOS) and National Centre for Computational Design and
  Discovery of Novel Materials (MARVEL), \'Ecole Polytechnique
  F\'ed\'erale de Lausanne, CH-1015 Lausanne, Switzerland}
\author{Giovanni Pizzi}
\email{giovanni.pizzi@epfl.ch}
\affiliation{Theory and Simulation of
  Materials (THEOS) and National Centre for Computational Design and
  Discovery of Novel Materials (MARVEL), \'Ecole Polytechnique
  F\'ed\'erale de Lausanne, CH-1015 Lausanne, Switzerland}

\date{\today}

\begin{abstract}

  We highlight with first-principles molecular dynamics the persistence of intrinsic $\langle111\rangle$ Ti off-centerings for BaTiO$_3$ in its cubic paraelectric phase. Intriguingly, these are inconsistent with the Pm$\bar 3$m space group often used to atomistically model this phase using density functional theory or similar methods.
  Therefore we deploy 
a systematic symmetry analysis to construct representative structural models in the form of supercells that satisfy a desired point symmetry but are built from the combination of lower-symmetry primitive cells. We define as structural prototypes the smallest of these that are both energetically and dynamically stable. Remarkably, two 40-atom prototypes can be identified for paraelectric BaTiO$_3$; these are also common to many other ABO$_3$ perovskites. These prototypes can offer structural models of paraelectric phases that can be used for the computational engineering of functional materials. Last, we show that the emergence of B-cation off-centerings and the primitive-cell 
phonon instabilities is controlled by the equilibrium volume, in turn dictated by the filler A cation.


\end{abstract}
\maketitle

Compounds with the perovskite structure are a versatile class of functional materials exhibiting a wide range of properties, such as 
superconductivity~\cite{Bednorz1988},
catalysis~\cite{Hwang2017}, 
photovoltaic energy harvesting~\cite{Jena2019} and ferroelectricity~\cite{Jona,Lines}.  
When ferroelectric, perovskites sustain a spontaneous polarization that can be switched with an electric field; as the temperature is raised, there is a transition above the Curie temperature to a paraelectric phase that has no net polarization.
Early studies of ~\bto~(a prototypical ABO$_3$ ferroelectric perovskite) suggested for these transitions a microscopic ``displacive'' model, in which local displacements of the B-cation (titanium) align with the macroscopic polarization~\cite{vonHippel1946,Merz1949}. 
For~\bto~this is along the~\one~direction in the rhombohedral ground state; as the temperature increases there is a transition to an orthorhombic phase above 183K, with the polarization along $\langle 110 \rangle$, then to a tetragonal phase above 278K, with the polarization along $\langle 100 \rangle$, before reaching the paraelectric cubic phase above 393K, with no net polarization~\cite{Merz1949}.
The results from diffuse X-ray scattering for all phases but the rhombohedral one~\cite{Comes1968, Comes1970} are somewhat inconsistent with such a displacive model.
This has led to the application of the
order-disorder model for the transitions~\cite{Bersuker1966,Chaves1976}
in which local polar displacements, driven by the pseudo Jahn-Teller effect~\cite{Bersuker2013}, are in different ordered arrangements in the ferroelectric phases at low temperatures, and become disordered in the paraelectric phase.
These two models can be reconciled if one considers 
the time-averaging inherent to most characterization techniques, which can effectively wash out the local displacements and present a higher-symmetry structure where the averaged displacements are aligned with the macroscopic polarization or cancel out~\cite{Stern2004}.

Microscopic displacements~\cite{Comes1968, Comes1970, Ravel1998, Zalar2003,Stern2004,Levin2014, Senn2016,Shi2018,Bencan2021} and phase transitions in perovskites have been studied extensively using effective Hamiltonians~\cite{Zhong1994, Zhong1995,Girshberg1999,Fu2003, Pirc2004,Walizer2006} or molecular dynamics, most often based on density-functional theory (DFT)~\cite{Krakauer1999, Ponomareva2008,Nishimatsu2008,Qi2016,Pasciak2018,Chen2020}.
Interestingly,~\bto~supercells possessing local~\one~Ti displacements and maintaining the experimentally-observed macroscopic polarization have been  shown to be energetically favorable~\cite{Zhang2006, Wang2020, Zhao2021} and dynamically stable~\cite{Zhang2006}, offering a unique insight into the microscopic potential-energy surfaces for these materials.

To elucidate the microscopic picture of paraelectricity in these perovskites, 
we performed Car--Parrinello molecular dynamics (CPMD) simulations of cubic~\bto, finding clear microscopic evidence of Ti off-centerings, and associated dipoles, along the~\one~directions which persist well above the Curie temperature, consistent with the order-disorder model.
\begin{figure}[tp]
\center
\includegraphics[width=\linewidth]{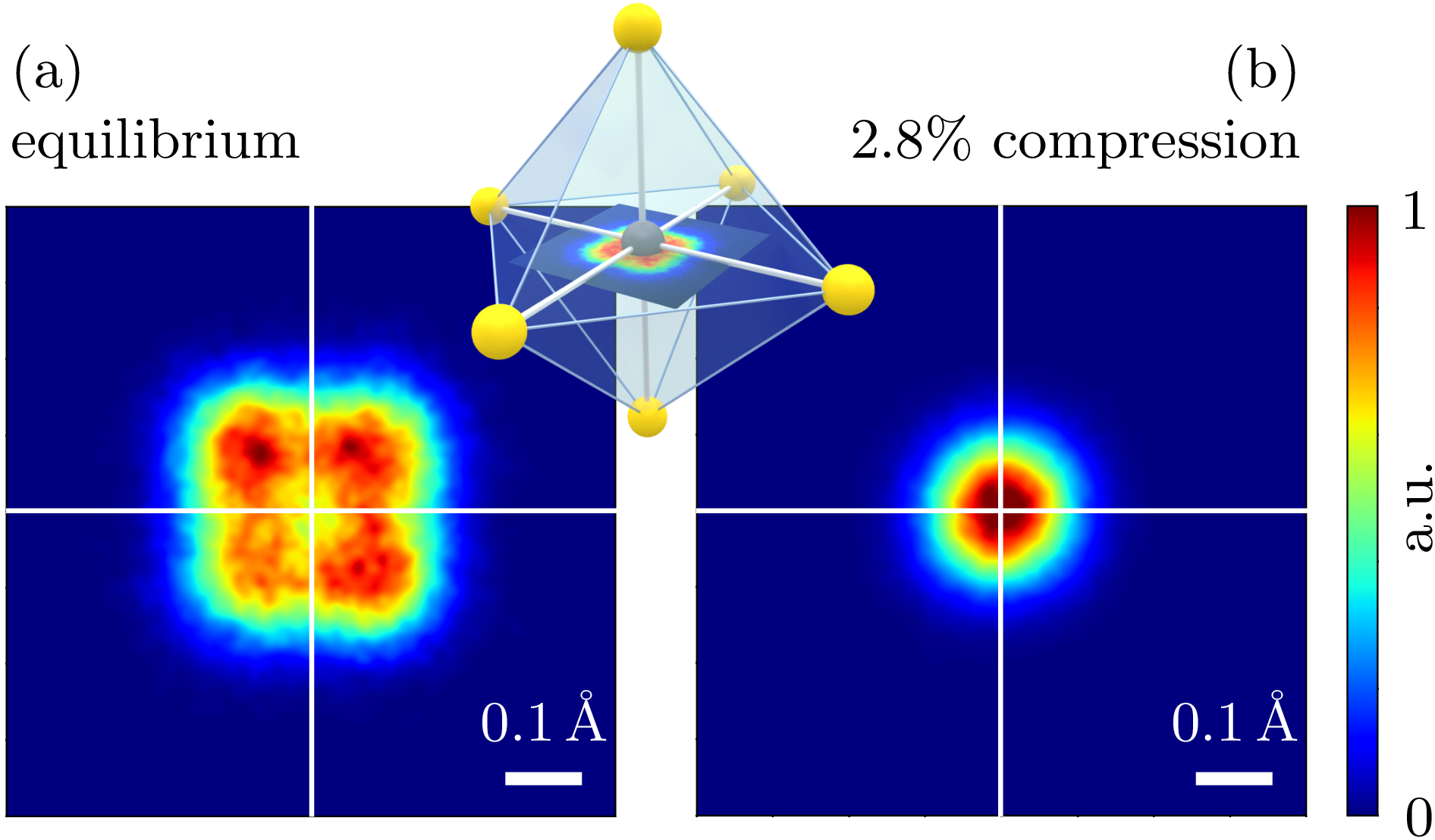}
\caption{
 CPMD simulation of ~\bto{}.
Histograms of the $xy$-plane projection of the displacement of Ti atoms with respect to the barycenter of the surrounding TiO$_6$ octahedron at (a) equilibrium volume (lattice parameter: 4.00~\AA) and (b) under a 2.8\% compressive hydrostatic strain (lattice parameter: 3.89~\AA), integrated over 13~ps of fixed-volume NVE CPMD simulations and on all Ti atoms of a 4$\times$4$\times$4 cubic supercell. The average temperature of the simulation is 315K. Inset: graphical 3D representation depicting the $xy$-plane projection of the Ti cation position in a representative TiO$_6$ octahedron.} 
\label{fig:MD}
\end{figure}
This is clearly apparent in Fig.~\ref{fig:MD}a, where we present the results of CPMD simulations at 315K for a 4$\times$4$\times$4 cubic supercell; the histogram for the Ti displacements projected onto one of the equivalent [100] planes shows how the Ti atoms always occupy off-center~\one~positions, rather than sitting at the center of their surrounding 
oxygen octahedron~\cite{note1}
(for the Methodology see SI Sec.~1~\cite{SIa}; for the associated data, see Ref. \onlinecite{MatCloud} on the Materials Cloud~\cite{Talirz2020}).
 We observe these off-center displacements up to temperatures around 450K; furthermore, they can be suppressed with compressive hydrostatic strain, resulting in an isotropic distribution (Fig.~\ref{fig:MD}b) in agreement with experimental measurements~\cite{Decker1989,Itie2006}. This observation will be relevant to the later discussion of volume effects and the role of A-site cations.

Inspired by these results, we aim here to systematically explore the microscopic structure of the paraelectric phase of ~\bto, to extend this exploration to other perovskites, and to lay the groundwork for a systematic analysis of phases that can possess ``hidden order'', including other ferroelectric or magnetic systems~\cite{Zhang2014,Yuan2019,Zhang2021} or those displaying higher-order couplings~\cite{Aeppli2020}. 
With this goal in mind, we introduce first the concept of microscopic templates, defined as lower-symmetry supercells that preserve a desired point symmetry (e.g., cubic). We then define microscopic prototypes as the smallest of these templates that are both energetically and dynamically stable (i.e., lower in energy, per formula unit, than the higher-symmetry primitive cell and with real, positive phonon dispersions),
thus minimizing computational cost while identifying the highest-symmetry, stable structures possessing the requisite symmetry. This approach is distinct yet complementary to that of special quasirandom structures (SQSs) when used to describe a polymorphous network, in which a single, large SQS exhibiting many local motifs is used. In this context, slightly different from the original development of SQSs to  characterize disordered alloys with first-principles calculation~\cite{Zunger1990, vandeWalle2013}, SQSs have been recently used to study paramagnetic phases~\cite{Kormann2012, Trimarchi2018} and complex perovskite-based systems~\cite{Lebedev2009,Voas2014,Varignon2019,Zhao2020, Zhao2021}.
In this work we develop instead a symmetry-based analysis and workflow, enumerating all possible supercells (up to a given size) with a desired point symmetry. In this way we identify not just local motifs, but more complex orderings which respect the desired global point symmetry. 
We describe it in the following and apply to structural microscopic prototypes,  but  these  concepts  can  be  equally  applied  to magnetic or electronic prototypes.

To identify structural prototypes we use group-subgroup relations, as discussed in Ref.~\onlinecite{Kozinsky2016}, to systematically enumerate all microscopic templates; here, we take the case of the cubic ABO$_3$ perovskite with space group Pm$\bar 3$m (international number 221), where the 1a, 1b and 3c Wyckoff positions are occupied by the A, B, and O atoms, respectively.
For each cubic subgroup of Pm$\bar 3$m, we define a cubic microscopic template as a supercell that can host symmetry-allowed displacements of A, B, and O atoms relative to their positions in the high-symmetry parent structure (group Pm$\bar 3$m) with no net polarization (see SI Sec.~1~\cite{SIa}
for further details).
Using 2$\times$2$\times$2 supercells  of 40 atoms,
we find 27 distinct cubic microscopic templates of group Pm$\bar 3$m, 10 of which host only oxygen displacements, while the remaining 17 allow the A and/or B cations to displace as well.
Table~\ref{table} summarizes the subgroups in which  the B cations can displace; 
see SI Sec.~2~\&~3~\cite{SIa}
for the complete list as well as a list duplicates that correspond to microscopic templates with higher symmetry.  
The same analysis can be applied to supercells of any desired size, but we find that in BaTiO$_3$ these 2$\times$2$\times$2  supercells are already sufficient to identify structural prototypes.

\begin{table}[tbp]
  \caption{
    The microscopic templates derived from the cubic subgroups (up to a 2$\times$2$\times$2  supercell) of parent group Pm$\bar 3$m (221) with at least one degree of freedom for the 1b Wyckoff position, allowing for B-site off-centering. 
    We list the subgroup (international short symbol, and number in parentheses) followed by the subgroup index (in square brackets);
    and the splittings of the three relevant Wyckoff positions (1a, 1b, 3c).
    These cells are a 2$\times$2$\times$2  supercell of the primitive cell with no translation.
  }
  \begin{tabular}{llclll}
    Group &&[Index] &
    1a & 1b & 3c \\ \hline\hline
Pm$\bar 3$m & (221)  &[8]
&  1a 1b 3c 3d &  8g &  12i 12j        \\
P$\bar 4$3n & (218) &[16]
&  2a 6b &  8e &  24i                  \\
I$\bar 4$3m & (217)  &[8]
&  2a 6b &  8c &  24g                  \\
P$\bar 4$3m & (215) &[16]
&  1a 1b 3c 3d &  4e 4e &  12i 12i     \\
Pa$\bar 3$  & (205) &[16]
&  4a 4b &  8c &  24d                   \\
Pm$\bar 3$  & (200) &[16]
&  1a 1b 3c 3d &  8i &  12j 12k        \\
I2$_1$3  & (199) &[16]
&  8a &  8a &  12b 12b                 \\
P2$_1$3  & (198) &[32]
&  4a 4a &  4a 4a &  12b 12b           \\
I23   & (197) &[16]
&  2a 6b &  8c &  24f                  \\
P23   & (195) &[32]
&  1a 1b 3c 3d  &  4e 4e &  12j 12j     \\
\end{tabular}
\label{table}
\end{table}

\begin{figure*}[t]
\includegraphics[width=\linewidth]{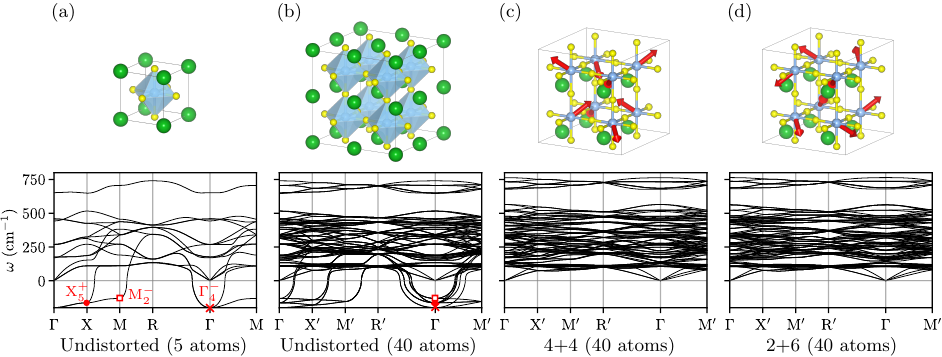}
\caption{Crystal structures (top) and phonon dispersions (bottom) for various BaTiO$_3$ cells, with Ba, Ti, and O atoms shown in green, blue and yellow, respectively.
(a) Five-atom primitive cubic cell (standard crystallographic labels from Ref.~\onlinecite{Hinuma2017}), displaying unstable phonon modes.
  The high-symmetry $\Gamma_4^-$, X$^+_5$ and M$^-_2$ instabilities are marked in red with a cross, a filled circle and an empty square, respectively.
(b) Forty-atom undistorted supercell; X$'$, M$'$ and R$'$ labels indicate the X, M and R points of the 2$\times$2$\times$2 supercell, respectively (the same labels are used also in panels c and d).
All three instabilities marked in panel a fold at $\Gamma$ in this supercell.
(c\&d) 40-atom supercell with the 4+4 and 2+6 displacement pattern, respectively.
In panels c and d (top), red arrows indicate the direction of atomic displacements (only shown for B cations for clarity).
We refer the reader to the SI Sec.~5~\cite{SIa} for the displacement pattern associated with each unstable mode and the other modes that contribute to the 4+4 and 2+6 displacement patterns, which can also be visualized with the Interactive phonon visualizer tool on the Materials cloud~\cite{Talirz2020}.
\label{fig:phonon}}
\end{figure*}

We then determine which of these microscopic templates, if any, are energetically stable.
Using variable-cell first-principles relaxations performed with Quantum ESPRESSO~\cite{Giannozzi2009,Giannozzi2017} using the PBEsol~\cite{Perdew2008} functional managed with AiiDA~\cite{Pizzi2016,Huber2020,Uhrin2021}, we take these 2$\times$2$\times$2 templates as starting structures and require that the point symmetry remains cubic (See SI Sec. 1~\cite{SIa} for further details). 
Remarkably, we find that two of the microscopic templates relax to supercells with non-trivial displacement patterns of the B cations; moreover, they display stable phonon dispersions across the entire Brillouin zone (see Fig.~\ref{fig:phonon}c,d). 
The remaining 25 templates either relax back to (the 2$\times$2$\times$2 supercell of) the five-atom primitive cell, well-known to be dynamically unstable (see Fig.~\ref{fig:phonon}a,b)~\cite{Ghosez1998,Ghosez1999} or to one of these two non-trivial displacement patterns. 
hese energetically and dynamically stable structures are the structural prototypes. As they are locally stable structures of the 0K potential energy landscape they serve as minimal models possessing the signature of the paraelectric phase, namely a global cubic symmetry but with local Ti displacements. These displacements, driven by local chemistry, can then also acquire correlations (e.g. linear chains~\cite{Comes1968}) that can be studied with large-scale molecular dynamics simulations~\cite{Ponomareva2008,Pasciak2010,Pasciak2018}.


The two structural prototypes have symmetry I$\bar 4$3m  and Pa$\bar 3$, respectively  (see Table~\ref{table}); their  structure and B-atom (Ti) displacement patterns  are shown in Fig.~\ref{fig:phonon}c-d.
We name these two prototypes 4+4 and 2+6 (for I$\bar 4$3m and Pa$\bar 3$, respectively), since 
considering any Ba atom, in the 4+4 (2+6) structure there are 4 (2) surrounding Ti atoms that displace toward it, while the remaining 4 (6) displace outwards.
We note that the 4+4 structure (I$\bar 4$3m) has been previously discussed in the work of Zhang \textit{et al.}~\cite{Zhang2006}.
The 4+4 and the 2+6 structural prototypes are lower in energy than the undistorted cubic structure by 11 and 15 meV/formula unit, respectively.  Furthermore, there is an energy barrier of only 3 meV/formula unit between these two structural prototypes (as found by nudged-elastic-band calculations, see SI Sec.~4~\cite{SIa}), suggesting that thermal fluctuations of the off-centerings do not require to go through the high-symmetry structure.

We contrast the phonon dispersions of the high-symmetry structure (Fig.~\ref{fig:phonon}a,b) with that of the 2 prototypes(Fig.~\ref{fig:phonon}c,d). The five-atom primitive cell displays instabilities at the zone-center $\Gamma$, belonging to the irreducible representation (irrep) $\Gamma_4^-$, and at the zone-boundary points X and M (irreps X$_5^+$ and M$_2^-$, respectively).
To gain further insight into the 4+4 and 2+6 patterns we analyze these with respect to the irreps of the five-atom-cell phonons using the ISODISTORT software ~\cite{isodistort,Campbell2006}.
We find that the displacements of both prototypes contain a mode with the symmetry of an unstable zone-boundary mode. 
Specifically,  the 4+4 prototype can be constructed by adding the displacements having the symmetry of the M$_2^-$ and M$_1^+$ irreps, while the 2+6 prototype originates from the X$_5^+$ and  M$_5^+$ irreps (see SI Fig. S2~\cite{SIa} for the  M$_1^+$ and M$_5^+$ modes).
Most importantly, out of the 27 distinct cubic templates, the 4+4 (I$\bar 4$3m) and 2+6 (Pa$\bar 3$) are the only 
ones with a displacement pattern that is constructed, in part, from a mode with the symmetry of an unstable mode of the parent structure, resulting in an appealing one-to-one correspondence between unstable zone-boundary phonon modes and prototypes with stable displacement patterns in 2$\times$2$\times2$ cubic supercells.
Wenote that the displacement patterns must occur in combination
with another mode in a cubic structure as they do not possess a
cubic point symmetry.
Furthermore, the $\Gamma_4^-$ mode is the polar instability and can only occur in lower-symmetry polar phases of BaTiO$_3$, which are therefore non-cubic.

\begin{figure}[tp]
\center
\includegraphics[width=\linewidth]{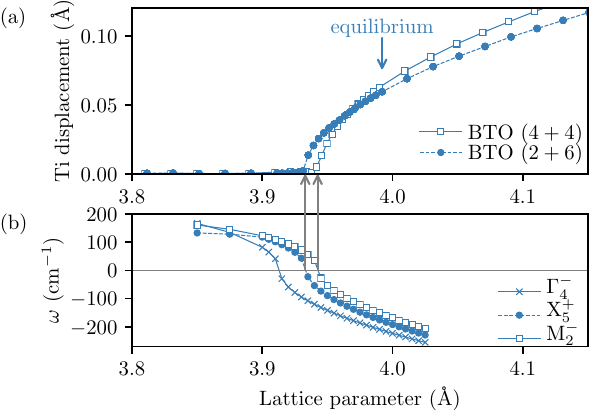}

\caption{
  (a) Magnitude of the Ti-atom displacements for the 4+4 and 2+6 patterns in \bto{} 
  as a function of the lattice parameter. Displacements are along the~\one~directions and we plot the displacement in angstrom along one Cartesian coordinate. 
The DFT (PBEsol) equilibrium lattice parameter in its lowest-energy cubic configuration is indicated by the blue arrow.
(b) Plot of the unstable phonon modes of BaTiO$_3$ with irrep $\Gamma_4^-$, X$_5^+$ and M$_2^-$ in the five-atom undistorted cubic cell as a function of the lattice parameter. The gray arrows indicate the lattice parameters at which the modes at X and M become unstable; in agreement with the corresponding displacement onsets in panel (a).
}
\label{fig:vol}
\end{figure}

We investigate in more detail 
the zone-boundary modes and the stability of the structural prototypes as a function of volume, prompted by the disappearance of the Ti off-centering under pressure in our CPMD simulations (Fig.~\ref{fig:MD}b) and in experiments~\cite{Decker1989,Itie2006}.
  We find that with increasing pressure the magnitude of the Ti displacements decreases for both  prototypes, and disappears when the lattice parameter is reduced by $\approx 1.6\%$, as reported in Fig.~\ref{fig:vol}a.
 We find that the Ti displacements as a function of volume can be fit by a double-well potential where the quadratic coefficient depends linearly on volume and changes sign at the onset of the displacements (see SI Sec.~6~\cite{SIa}). 
 This suggests that at least one phonon mode associated with this structural prototype becomes unstable at the same volume where the Ti displacement becomes energetically favorable. In Fig.~\ref{fig:vol}b we plot as a function of the lattice parameter the phonon frequencies for the $q=\Gamma,$ X and M modes that are unstable in the five-atom primitive cell (irreps $\Gamma_4^-$, X$_5^+$ and M$_2^-$, respectively).
We find that expanding the volume further softens these modes, while applying pressure stabilizes them, in agreement with previous calculations~\cite{Cohen1990}.
The fact that the $\Gamma_4^-$ mode also stabilizes at a lower lattice parameter is indicative of a pressure at which the system could be ferroelectric below a critical temperature, but no Ti displacements would be observed in the cubic paraelectric phase.
Notably, the M$_2^-$ and X$_5^+$ modes become unstable at the same lattice parameter where the 4+4 and 2+6 displacement patterns respectively emerge (gray arrows in Fig.~\ref{fig:vol}b).

\begin{figure}[t!]
  \center
    \includegraphics[width=\linewidth]{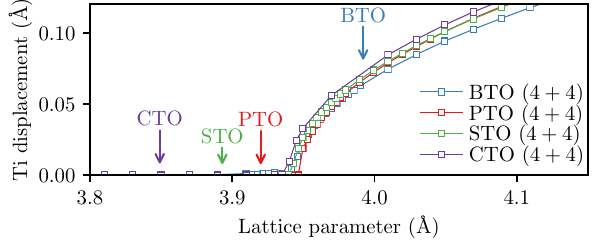}
    \caption{
      Magnitude of the Ti-atom displacements for the 4+4 pattern in (Ba,Pb,Sr,Ca)TiO$_3$ (labeled as BTO, PTO, STO and CTO) as a function of the lattice parameter. Displacements are along the~\one~directions and we plot the displacement in angstrom along one Cartesian coordinate.
The DFT (PBEsol) equilibrium lattice parameter is indicated by the arrow of the corresponding color. A clear universal trend across the titanates is demonstrated; the 4+4 pattern is stable only in unstrained BaTiO$_3$  because of the larger A cation and thus larger lattice parameter. }
  \label{fig:disp44}
\end{figure}

Thus, the 4+4 and 2+6 prototypes originate from the unstable M$_2^-$ and X$_5^+$ modes, which do not involve A-cation displacements. To test the effect of the A cation we extend the study to  PbTiO$_3$, SrTiO$_3$, and CaTiO$_3$. 
We report in Fig.~\ref{fig:disp44} the results for the 4+4 prototype, highlighting a universal trend where the B-site displacement as a function of lattice parameter is broadly 
independent of the chosen A cation.
The stability of the prototype, and thus the nature of the paraelectric phase, is instead determined by the equilibrium lattice parameter  – indicated in Fig.~\ref{fig:disp44} by arrows – which is largely determined by the A cation. For Pb, Sr, and CaTiO$_3$ the lattice parameter is smaller than the critical value at which the displacement pattern becomes energetically favorable, $\sim$3.95\AA.  
For all titanates studied, the displacement pattern onset occurs at the lattice parameter at which the M$_2^-$ mode becomes unstable. A similar picture emerges for the 2+6 pattern (except for CaTiO$_3$, due to its significantly smaller lattice parameter) -- see SI Sec.~7~\cite{SIa}.

Testing a broader range of 49 perovskites from Ref.~\onlinecite{Armiento2014} shows that B-site off-centerings along~\one~directions provides prototypes at the relaxed equilibrium lattice parameter not only for~\bto, but for most zirconates, niobates, and tantalates, CaHfO$_3$ and BiScO$_3$, as reported in SI Sec.~8~\cite{SIa}. 
However, the energetic stability of these prototypes as a function of lattice parameter is B-site specific. This is expanded on in SI Sec.~9~\cite{SIa} 
where we further investigate the stability of the 4+4 and 2+6 displacement patterns as a function of lattice parameter in the titanates, niobates and zirconates, demonstrating the universality of the occurrence of B-cation displacements, and their strong, family-specific volume dependence.

The relationship we have observed between the unstable zone-boundary phonons of the primitive cubic structure and the  displacement patterns as a function of lattice parameter indicates that, at a given volume, one could use the unstable phonon modes to predict which microscopic templates would result in 
structural prototypes.
To verify the robustness of our conclusions against the choice of DFT functional, we tested for~\bto~the dependence on the functional, finding that the Ti displacement amplitudes are independent of the functional choice (see SI Sec.~10~\cite{SIa}
), but that the functional determines the equilibrium volume. This highlights the need to choose a functional that accurately reproduces the experimental lattice parameter in order to correctly predict which prototypes occur at the equilibrium volume.

 In summary, motivated by the observed persistence of~\one~Ti off-centerings above the Curie temperature in our CPMD simulations of~\bto, 
 we  systematically identify  microscopic  structural  prototypes of the paraelectric phase, \textit{i.e.}, the smallest supercells with cubic point symmetry that are 
 simultaneously energetically and dynamically stable.
  These cubic prototypes, hosting stable local dipoles due to the~\one~Ti displacements in a cubic paraelectric phase, are found through a symmetry analysis exploring all possible 40-atom microscopic templates, followed by density-functional theory and density-functional perturbation theory calculations to assess energetic and dynamical stability.
 Moreover, we highlight how off-centering amplitudes are strongly dependent on volume, and relate their patterns to the zone-boundary unstable  phonons of the five-atom undistorted primitive cubic cell, suggesting a predictor for the identification of such prototypes.
 These cubic prototypes would be challenging to identify without the present symmetry-based approach, due to the combinatorial complexity of large supercells and the attractive basin of the rhombohedral five-atom ground state associated with the polar instability.
 We highlight that these prototypes can serve as minimal models of the paraelectric phase in first-principles calculations of response functions with the correct tensorial symmetry as they provide a faithful microscopic representation with key features: the persistence of local Ti displacements and the appropriate macroscopic point group.

We finally emphasize that this approach is general. 
Beyond its extension to study the  prevalence of the B-site~\one~off-centerings in ABO$_3$ perovskites, it can be used in any crystalline system 
to find candidate templates and efficiently search for prototypes that are local minima in the potential-energy surface, providing an in-depth study of, for example, the electronic or magnetic properties of a polymorphic system. 
This approach lays the foundation to investigate dynamics, thermodynamics and chemical substitutions, as these prototypes could be used to capture subtle details of the energy landscape and to provide models to study the properties or transitions of disordered phases, such as alloys, paramagnetic phases, or defects in paraelectric phases.

\begin{acknowledgments}
We greatly acknowledge A. Cepellotti and K. M. Rabe for useful discussions. M.K.\@ and N.M.\@ acknowledge funding from the Samsung Advanced Institute of Technology; G.P.\@ and N.M.\@ from the MARVEL NCCR, a National Centre of
Competence in Research, funded by the Swiss National Science Foundation (Grant No. 182892) and from the European Centre of Excellence MaX ``Materials design at the Exascale'' (824143); B.K. from Robert Bosch LLC; computational support has been provided by the 
Swiss National Supercomputing Centre CSCS under project ID s1073.  All data required to reproduce this work is available at Ref.~\onlinecite{MatCloud} in the Materials Cloud.
\end{acknowledgments}
\vspace{-10pt}
\putbib
\end{bibunit}

\begin{bibunit}
\clearpage

\onecolumngrid

\appendix  

\renewcommand{\thesection}{\arabic{section}} 
\renewcommand{\thesubsection}{\Roman{subsection}} 
\setcounter{figure}{0}
\setcounter{table}{0}
\renewcommand{\thetable}{S\arabic{table}}
\renewcommand{\thefigure}{S\arabic{figure}}
\centering{\bf Supplemental Material: Microscopic picture of paraelectric perovskites}

 \flushleft

\section{Methods}
\label{SI:meth}
Our first-principles DFT-based calculations are performed using Quantum ESPRESSO~\cite{Giannozzi2009,Giannozzi2017} with the PBEsol~\cite{Perdew2008} functional, RRKJ pseudopotentials~\cite{Rappe1990} and wavefunction and charge-density energy cutoffs of 60 and 600 Ry, respectively.

\subsection{Car-Parrinello Molecular Dynamics}
CPMD calculations~\cite{Car1985} are carried out using the \texttt{cp.x} code of Quantum ESPRESSO in the NVE ensemble (constant number of particles, volume and energy) on a $4 \times 4 \times 4$ supercell (320 atoms) using a $\Gamma$-only sampling, where the initial atomic positions are chosen so that, after an initial equilibration (1.5~ps), the average temperature is the one reported in the figure caption. The lattice parameter is also increased with respect to the $T=0$~K value so as to account for the experimentally-measured thermal expansion~\cite{Bland1959}.
The wavefunction and charge-density energy cutoffs is set to 40 and 320 Ry, respectively. A time step of 0.25~fs
and an electron mass of 400~$m_e$ are used.
\subsection{Symmetry analysis to determine the microscopic templates}

We search for all the subgroups with a desired point symmetry (here, cubic) of the high-symmetry space group with unique Wyckoff splittings in a recursive fashion. 
We use and compare both the ISOTROPY command-line tool~\cite{iso} and the tools available on the Bilbao Crystallographic server~\cite{bilbao,Aroyo2006}, in particular the CELLSUB program to obtain the list of subgroups, the conjugacy classes and the transformation matrices, then used as input to the WYCKSPLIT program~\cite{Kroumova1998} to get the splitting of the relevant Wyckoff positions. 
These two set of rules yield the same results and allow us to generate the group--subgroup relationship used to build Tables~1 (main text), \ref{table-27} and \ref{table-trivial}. 
To confirm the space group, removing the duplicate microscopic templates, we perform a final check with spglib~\cite{Togo2018}.

\subsection{Density functional theory calculations to search for structural prototypes}
Using a random displacement on the order of 0.05~\AA{} for each Wyckoff-position free parameter, we construct cells for each of the 27 cubic microscopic templates associated with each unique subgroup for the cubic phase of \bto~and perform variable-cell relaxations constraining cubic point symmetry.
At every ionic minimization, the forces were symmetrized, effectively enforcing symmetries detected at the onset of the calculation. 
This ensures that the system reaches the minimum-energy structure with the constraint of possessing the symmetry of the parent space group (e.g., a given cubic spacegroup), avoiding that the system relaxes to one of the lower-symmetry structures (e.g., the rhombohedral structure that is the ground state at $T=0~K$ for BaTiO$_3$).
This enforces the cubic point symmetry possessed by all of the 27 microscopic templates.
The calculations are managed using the AiiDA framework~\cite{Pizzi2016,Huber2020}, a high-throughput platform that allows to automatically launch, retrieve, parse and organize the calculations, storing the results in a database and automatically managing sequences of calculations via its workflow engine.
Given a microscopic template as a starting structure, we relax both the lattice parameters and the internal coordinates with a force tolerance of 5$\times$10$^{-5}$ eV/\AA{} 
and an energy tolerance of of 3$\times$10$^{-11}$ eV 
using a $ 3 \times 3 \times3$ $k$-mesh in the $2\times2\times 2$ supercells (and equivalent meshes in different cell sizes, e.g., $6 \times 6 \times6$ in the 5-atom unit cell).

To calculate the potential energy landscape, we perform a nudged-elastic-band (NEB) calculation~\cite{Henkelman2000} using the metastable 4+4 and 2+6 structures and the undistorted structure as the constrained points.
To calculate the phonon dispersion we use density functional perturbation theory (DFPT)~\cite{Giannozzi1991,Baroni2001} as implemented in Quantum ESPRESSO with a $2\times 2\times 2$ $q$-mesh imposing the acoustic sum rule. For the phonon dispersions, 50 points are used along each segment for the Fourier interpolation.
We use ISODISTORT~\cite{isodistort,Campbell2006} from the ISOTROPY package to analyze the symmetry of the unstable modes and of the displacement patterns of the metastable lower-symmetry structures, i.e. the structural prototypes.

\clearpage

\section{27 distinct subgroups of space group 221 \\ for occupied Wyckoff positions 1a, 1b, 3c}
\label{SI:table27}
\begin{table}[hbp]
  \caption{
    The 27 distinct cubic subgroups of parent group Pm$\bar 3$m (221) possessing  non-trivial splittings of the 1a, 1b and 3c Wyckoff positions are reported which each correspond to a microscopic template, including the 10 subgroups reported in the main text.
    We list the subgroup (international short symbol and number) followed by the subgroup index (in square brackets); the transformation matrix $T_x$; and the splittings of the three relevant Wyckoff positions.
    $T_x$ is defined as the transformation from the primitive cell to the $2 \times 2 \times 2$ cell (with matrix equal to twice the identity), and with an origin shift $(x,x,x)$. In the first ten subgroups (above the dividing line), only oxygen atoms can displace from the high-symmetry structure. 
    \label{table-27}
}

\begin{tabular}{llcllll}
Group &&[Index] &
$T_x$ & 1a & 1b & 3c \\ \hline\hline
Im$\bar 3$m & (229)  &[4]
& $T_{1/2}$ &  8c    &  2a 6b &  12e 12d \\ 
Im$\bar 3$m & (229)  &[4] 
& $T_{0}$&  2a 6b &  8c    &  24h     \\ 
Fm$\bar 3$m & (225)  &[2]
& $T_{1/2}$&  8c    &  4a 4b &  24e     \\ 
Pn$\bar 3$m & (224)  &[8]
& $T_{1/2}$&  2a 6d &  4b 4c &  24k     \\ 
Pm$\bar 3$n & (223)  &[8]
& $T_{0}$&  2a 6b &  8e    &  24k     \\ 
P4$_2$32 & (208) &[16]
& $T_{0}$&  2a 6d &  4c 4b &  24m     \\ 
Ia$\bar 3$  & (206)  &[8]
& $T_{0}$&  8a    &  8b    &  24d     \\ 
Im$\bar 3$  & (204)  &[8]
& $T_{0}$&  2a 6b &  8c    &  24g     \\ 
Im$\bar 3$  & (204)  &[8]
& $T_{1/2}$&  8c    &  2a 6b &  12d 12e \\ 
Pn$\bar 3$  & (201) &[16]
& $T_{1/2}$&  2a 6d &  4b 4c &  24h     \\ 
\hline           
Pm$\bar 3$m & (221)  &[8]
& $T_{1/2}$&  8g &  1a 3c 3d 1b &  6e 12h 6f      \\
Pm$\bar 3$m & (221)  &[8]
& $T_{0}$&  1a 3d 3c 1b &  8g &  12i 12j        \\
P$\bar 4$3n & (218) &[16]
& $T_{0}$&  2a 6b &  8e &  24i                  \\
I$\bar 4$3m & (217)  &[8]
& $T_{0}$&  2a 6b &  8c &  24g                  \\
I$\bar 4$3m & (217)  &[8]
& $T_{1/2}$&  8c &  2a 6b &  12e 12d              \\
P$\bar 4$3m & (215) &[16]
& $T_{1/2}$&  4e 4e &  1a 3c 3d 1b &  6f 12h 6g   \\
P$\bar 4$3m & (215) &[16]
& $T_{0}$&  1b 3c 3d 1a &  4e 4e &  12i 12i     \\
Pa$\bar 3$  & (205) &[16]
& $T_{1/2}$&  8c &  4a 4b &  24d                  \\
Pa$\bar 3$  & (205) &[16]
& $T_{0}$&  4a 4b &  8c &  24d                  \\
Pm$\bar 3$  & (200) &[16]
& $T_{1/2}$&  8i &  1a 3c 3d 1b &  6e 6g 6f 6h    \\
Pm$\bar 3$  & (200) &[16]
& $T_{0}$&  1a 3d 3c 1b &  8i &  12j 12k        \\
I2$_1$3  & (199) &[16]
& $T_{0}$&  8a &  8a &  12b 12b                 \\
P2$_1$3  & (198) &[32]
& $T_{0}$&  4a 4a &  4a 4a &  12b 12b           \\
I23   & (197) &[16]
& $T_{0}$&  2a 6b &  8c &  24f                  \\
I23   & (197) &[16]
& $T_{1/2}$&  8c &  2a 6b &  12d 12e              \\
P23   & (195) &[32]
& $T_{1/2}$&  4e 4e &  1a 3c 3d 1b &  6f 6h 6g 6i \\
P23   & (195) &[32]
& $T_{0}$&  1a 3d 3c 1b &  4e 4e &  12j 12j     \\
\end{tabular}
\end{table}

\clearpage

\section{Duplicate and trivial subgroups of space group 221 \\ for occupied 
Wyckoff positions 1a, 1b, 3c}
\label{SI:tablesdup}

\begin{table}[h]
  \caption{
    In addition to the 27 cubic subgroups of parent group Pm$\bar 3$m (221) reported in the main text, there are 10 cubic subgroups with trivial splittings of the 1a, 1b and 3c Wyckoff positions and 15 duplicate subgroups.
In the trivial subgroups, the resulting split Wyckoff positions have no degrees of freedom. The cell of the representation of the first four subgroups is the primitive cell and of the remaining six is the $2 \times 2 \times 2$ cell. 
    The duplicate subgroups, identified using spglib~\cite{Togo2018}, when only the Wyckoff positions originating from the 1a, 1b and 3c positions in the parent group are occupied, actually have higher symmetry and fall back in one of the subgroups of Table~\ref{table-27}, reported in the right-most columns.
   For these duplicate subgroups, there are 12 in which only the oxygens have degrees of freedom, and 3 in which the A and/or B sites have degrees of freedom. 
    We report the subgroup using the international short symbol; the subgroup number followed by the subgroup index (in square brackets); the transformation matrix $T_x$; and the splittings of the three relevant Wyckoff positions.
    $T_x$ is defined as the transformation from the primitive cell to the $2 \times 2 \times 2$ cell (with matrix equal to twice the identity), and with an origin shift $(x,x,x)$. For the four subgroups with primitive cell representations, the transformation matrix is trivial.
   The reported subgroup index and transformation matrix, in all cases, is given with respect to the parent group Pm$\bar 3$m (221) with the 1a, 1b, and 3c positions occupied. 
    \label{table-trivial}}
\begin{tabular}{llcllllllcl}
Group &&[Index] &
$T_x$ & 1a & 1b & 3c &
Equivalent &group&[Index]&$T_x$\\ \hline\hline
\multicolumn{11}{c}{Trivial subgroups} \\ \hline
P$\bar 4$3m & (215)  &[2]
& &  1a    &  1b &  3c & Pm$\bar 3$m & (221) &[1] &\\
P432 & (207)  &[2]
& &  1a    &  1b &  3c & Pm$\bar 3$m & (221)  &[1] &\\
Pm$\bar 3$ & (200)  &[2]
&  &  1a    &  1b &  3c & Pm$\bar 3$m & (221)  &[1]& \\
P23& (195)  &[4]
&  &  1a    &  1b &  3c & Pm$\bar 3$m & (221)  &[1] &\\
Fm$\bar 3$c & (226)  &[2]
& $T_{0}$ &  8a    &  8b &  24d & Pm$\bar 3$m & (221)  &[1] &\\
Fm$\bar 3$c & (226)  &[2]
& $T_{1/2}$ &  8b    &  8a &  24c & Pm$\bar 3$m & (221)  &[1] &\\
Fm$\bar 3$m & (225)  &[2]
& $T_{0}$ &  4a 4b    &  8c &  24d & Pm$\bar 3$m & (221)  &[1] &\\
F$\bar 4$3c & (219)  &[4]
& $T_{0}$ &  8a    &  8b &  24c & Pm$\bar 3$m & (221) &[1] &\\
F432 & (209)  &[4]
& $T_{0}$ &  4a 4b   &  8c &  24d & Pm$\bar 3$m & (221)  &[1] &\\
Fm$\bar 3$ & (202)  &[4]
& $T_{0}$ &  4a 4b    &  3c &  24d & Pm$\bar 3$m & (221)  &[1] &\\
\hline
\multicolumn{11}{c}{Duplicate subgroups, only oxygen degrees of freedom} \\ \hline
Pn$\bar 3$m & (224)  &[8]& $T_{1/2}$&  4b 4c &  2a 6b &  12g 12h & Im$\bar 3$m & (229) &[4] &$T_{1/2}$\\
Pm$\bar 3$n & (223)  &[8]& $T_{1/2}$&  8e &  2a 6b &  6c 6d 12f & Im$\bar 3$m & (229)  &[4] &$T_{1/2}$\\
Pn$\bar 3$n & (222)  &[8]& $T_{0}$&  8c &  2a 6b &  12e 12d & Im$\bar 3$m & (229)  &[4] &$T_{1/2}$\\
Pn$\bar 3$n & (222)  &[8]& $T_{1/2}$&   2a 6b &  8c & 12e 12d & Im$\bar 3$m & (229)  &[4] &$T_{0}$\\
F$\bar 4$3m & (216)  &[4]& $T_{0}$&  4a 4b &  4c 4d &  24g & Fm$\bar 3$m & (225)  &[2] &$T_{1/2}$\\
I432 & (211)  &[8]& $T_{1/2}$&  8c &  2a 6b &  12e 12d & Im$\bar 3$m & (229)  &[4] &$T_{0}$\\
I432 & (211)  &[8]& $T_{0}$ &  2a 6b &  8c&  24h & Im$\bar 3$m & (229)  &[4] &$T_{1/2}$\\
F432 & (209)  &[4]& $T_{0}$ &  2a 6b &  8c&  24e & Fm$\bar 3$m & (225)  &[2] &$T_{1/2}$\\
P4$_2$32 & (208)  &[16]&$T_{1/2}$&  4b 4c &  2a 6d &  6e 6f 12d& Im$\bar 3$m & (229)  &[4]& $T_{1/2}$\\
Fm$\bar 3$ & (202)  &[4]& $T_{1/2}$&  8c &  4a 4b &  24e & Fm$\bar 3$m & (225)  &[2] &$T_{1/2}$\\
Pn$\bar 3$ & (201)  &[16]& $T_{0}$&  4b 4c & 2a 6d&   12f 12g & Im$\bar 3$ & (204)  &[8]& $T_{1/2}$ \\
F23 & (196)  &[8]& $T_{0}$&  4a 4b &  4c 4d &  24g & Fm$\bar 3$m & (225)  &[2] &$T_{1/2}$\\
\hline
\multicolumn{11}{c}{Duplicate subgroups, A and/or B site degrees of freedom} \\ \hline
P$\bar 4$3n & (218)  &[16]& $T_{1/2}$&  8e & 2a 6b&   6c 6d 12f & I$\bar 4$3m & (217)  &[8] &$T_{1/2}$\\
P432 & (207)  &[16]& $T_{1/2}$&  8g & 1a 1b 3c 3d &   6e 6f 12h & Pm$\bar 3$m & (221)  &[8] &$T_{1/2}$\\
P432 & (207)  &[16]& $T_{0}$&  1a 1b 3c 3d & 8g&   12i 12j & Pm$\bar 3$m & (221)  &[8] &$T_{0}$\\
\end{tabular}
\end{table}

\clearpage

\section{Energy barriers}
  \label{SI:NEB}
\begin{figure}[h]
  \centering
  \includegraphics[width=.6\linewidth]{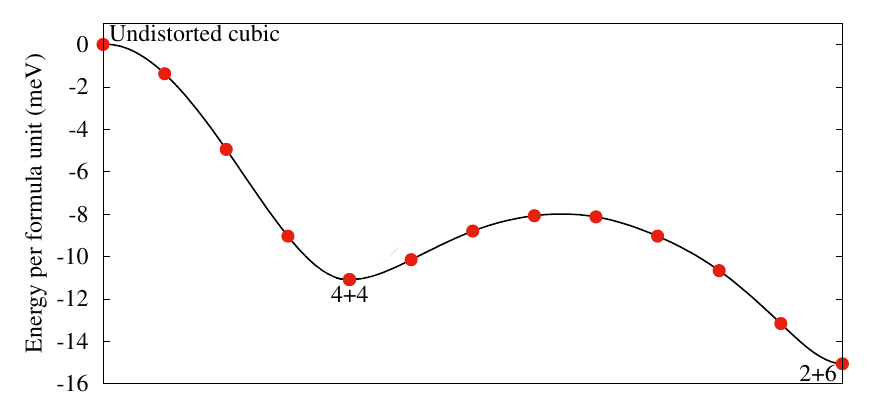}
  \caption{Potential energy landscape calculated via two nudged-elastic-band calculations between the undistorted cubic \bto{} (used as the reference energy) and the relaxed 4+4 structure, and between the relaxed 4+4 and 2+6 structures.
    The calculations show that the barrier between the 4+4 and 2+6 structures is lower than the energy difference with respect to the undistorted cubic \bto.
We note that the non-centrosymmetric lower-temperature phases of BaTiO$_3$, associated with the polar instability $\Gamma_4^-$ are lower in energy that the 4+4 and 2+6 structural prototypes.}
\end{figure} 

\section{Unstable B\lowercase{a}T\lowercase{i}O$_3$ modes and decomposition of the stable prototypes}
  \label{SI:modes}
\begin{figure}[h]
  \centering
  \includegraphics[width=.6\linewidth]{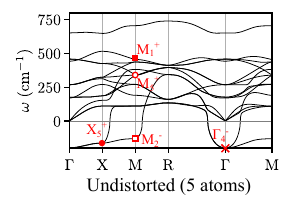}
  \caption{ Phonon dispersion of the 5-atom primitive cubic cell, displaying unstable phonon modes.
The high-symmetry $\Gamma_4^-$, X$^+_5$ and M$^-_2$ instabilities are marked in red with a cross, a filled circle and an empty square, respectively. Furthermore the M$_1^+$ and M$_5^+$ modes which occur in the 4+4 and 2+6 stable prototypes, respectively, are marked with a red filled square and an open circle.
}
\end{figure} 
\clearpage
\begin{figure}[h!] 
    \centering
    \includegraphics[width=.8\linewidth]{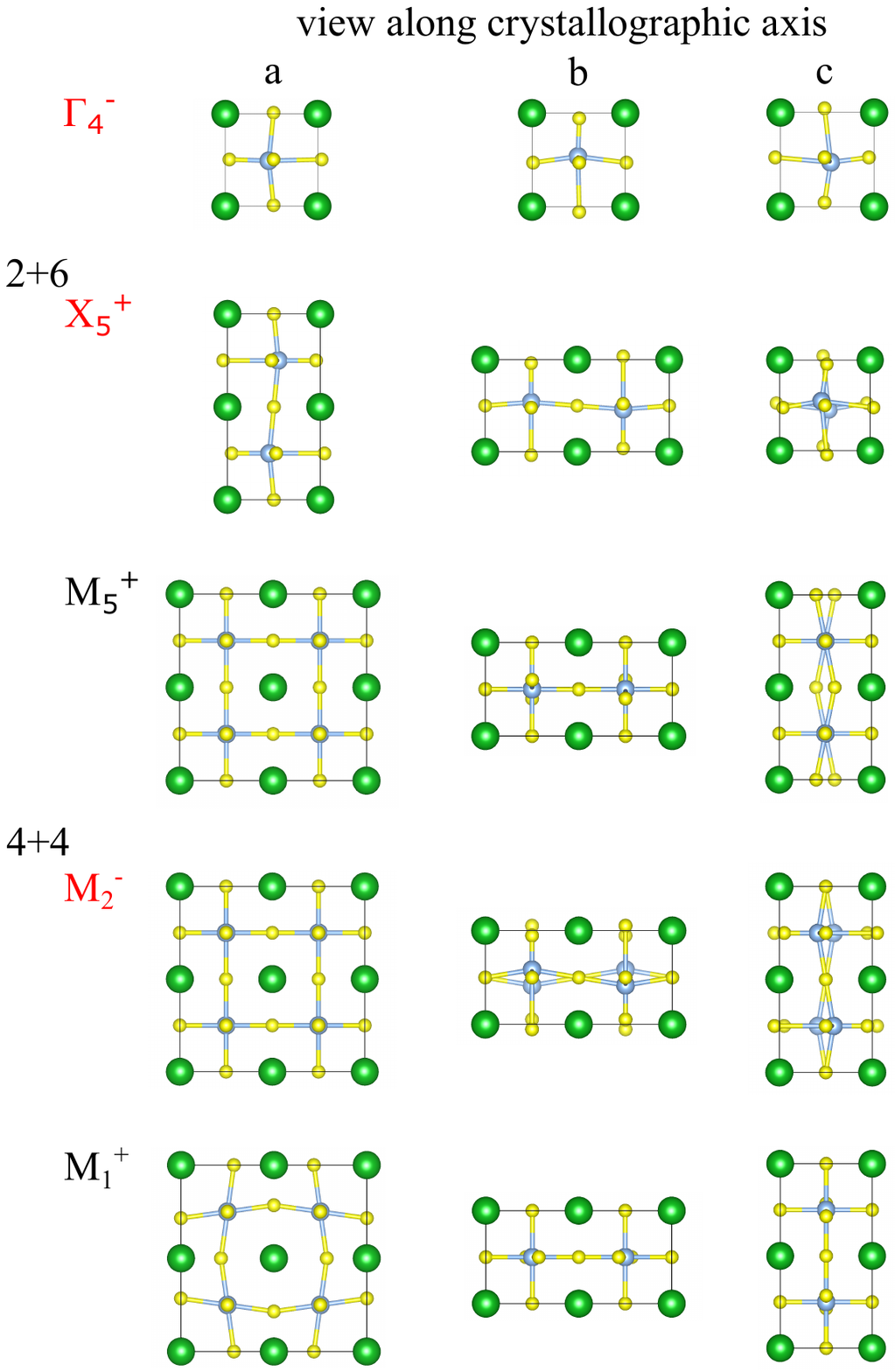}
    \caption{
    The displacement patterns associated with the unstable  $\Gamma_4^-$, M$_2^-$,  and X$_5^+$ modes (labelled in red) as well as the M$_1^+$ and M$_5^+$ modes which occur in the 4+4 and 2+6 stable prototypes, respectively - each labelled in the preceeding figure. Each mode is shown in its own primitive cell. We note that the  $\Gamma_4^-$, X$_5^+$, and M$_5^+$ modes are all two-fold degenerate.
    All data (inputs and outputs) from our self-consistent DFT calculations as well as the DFPT phonon calculations are available in a Materials Cloud Archive entry at the following DOI: \href{https://archive.materialscloud.org/record/2021.104}{https://doi.org/10.24435/materialscloud:pg-50}, within the file \texttt{materials\_5atom\_phonon.zip} file, in the directory \texttt{materials\_5atom\_phonon/BaTiO3/3.98}; the displacement patterns associated with these phonon modes can be visualized in an interactive manner, using the ``Interactive phonon visualizer'' tool on the Materials Cloud, available at the following address: \href{https://www.materialscloud.org/work/tools/interactivephonon}{https://www.materialscloud.org/work/tools/interactivephonon}.
    }
    \label{fig:response}
\end{figure}
\clearpage

\section{Double-well model for displacement pattern onset in Barium Titanate}
 \label{SI:vol}
\vspace{-10pt}
 Using a simple double-well model we can fit the Ti displacement as a function of volume. Assuming a simple double-well potential where the central point is the undistorted cubic structure and the minima of the wells correspond to the 4+4 (or 2+6) metastable structure, we define
 \begin{align}
E(x) = A(V)x^2+B(V) x^4, \label{Ex}
 \end{align}
 where $E$ is the total energy (with the zero set at the energy of the undistorted structure), $x$ is the set of coordinates that takes the undistorted structure to the metastable state (we will use here the magnitude of Ti displacements), and the coefficients $A$ and $B$ depend on the volume of the system. Solving for the stationary points of Eq.~\eqref{Ex}, we can find the minima at $(\tilde x(V), \tilde E(V))$, and then invert the solutions to obtain:
 \begin{align}
   A(V) = 2\tilde E/\tilde x^{2}, \qquad B(V) = -\tilde E/\tilde x^{4}. \label{AB}
 \end{align}

\begin{figure}[h]
\centering
\begin{minipage}{.45\textwidth}
  \centering
  \includegraphics[width=0.93\linewidth]{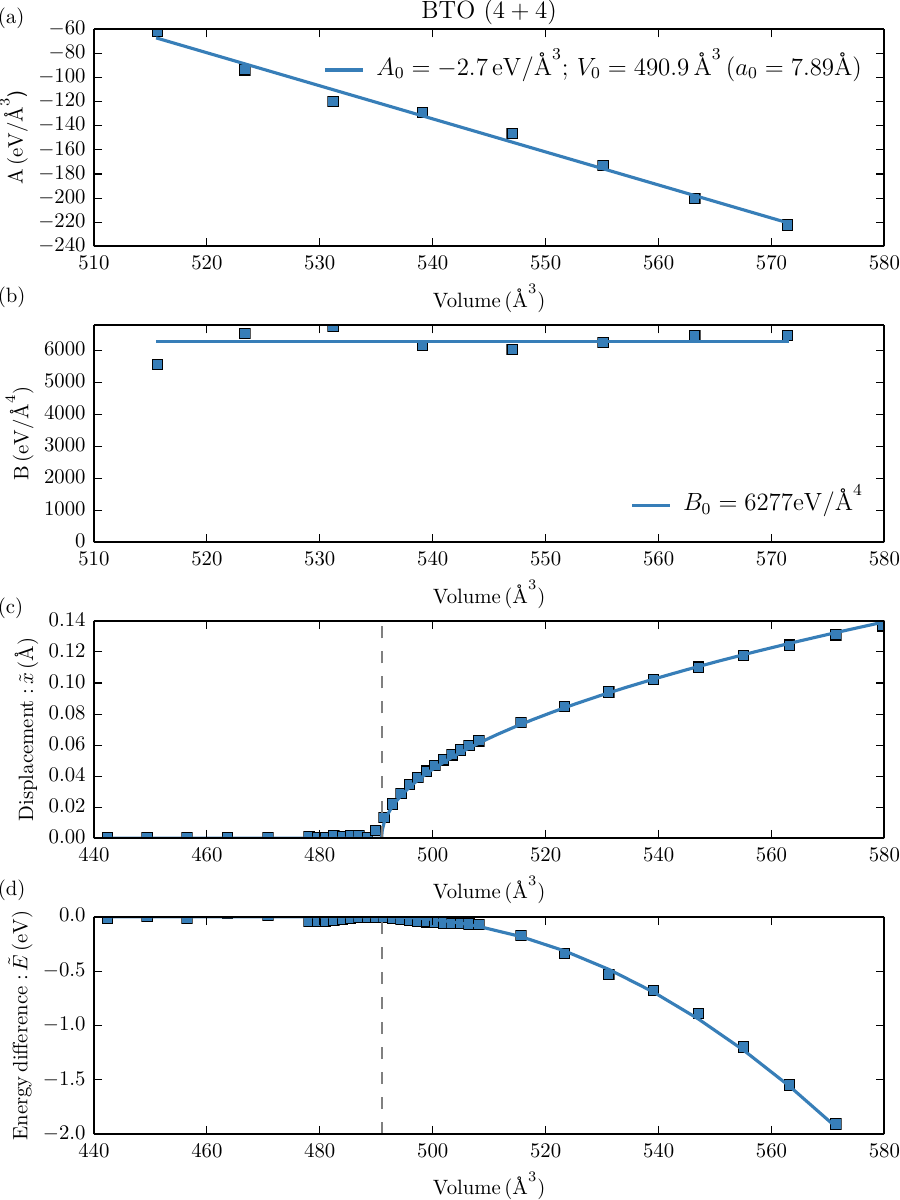}
  \caption{Double-well model for the 4+4 displacement pattern.
  (a) Fit of $A(V)=A_0\cdot (V-V_0)$; (b) fit of $B(V)=B_0$; (c) displacement $\tilde{x}(V)$ and analytical expression using the fit parameters from panels a and b; and, (d) energy difference $\tilde{E}(V)$ of the relaxed displacement patterns w.r.t. the undistorted structure.
  }
  \label{fig:dw4-4}
\end{minipage}%
\begin{minipage}{.05\textwidth}
  \centering
  \color{white}{d}
 \end{minipage}%
\begin{minipage}{.45\textwidth}
  \centering
  \includegraphics[width=0.93\linewidth]{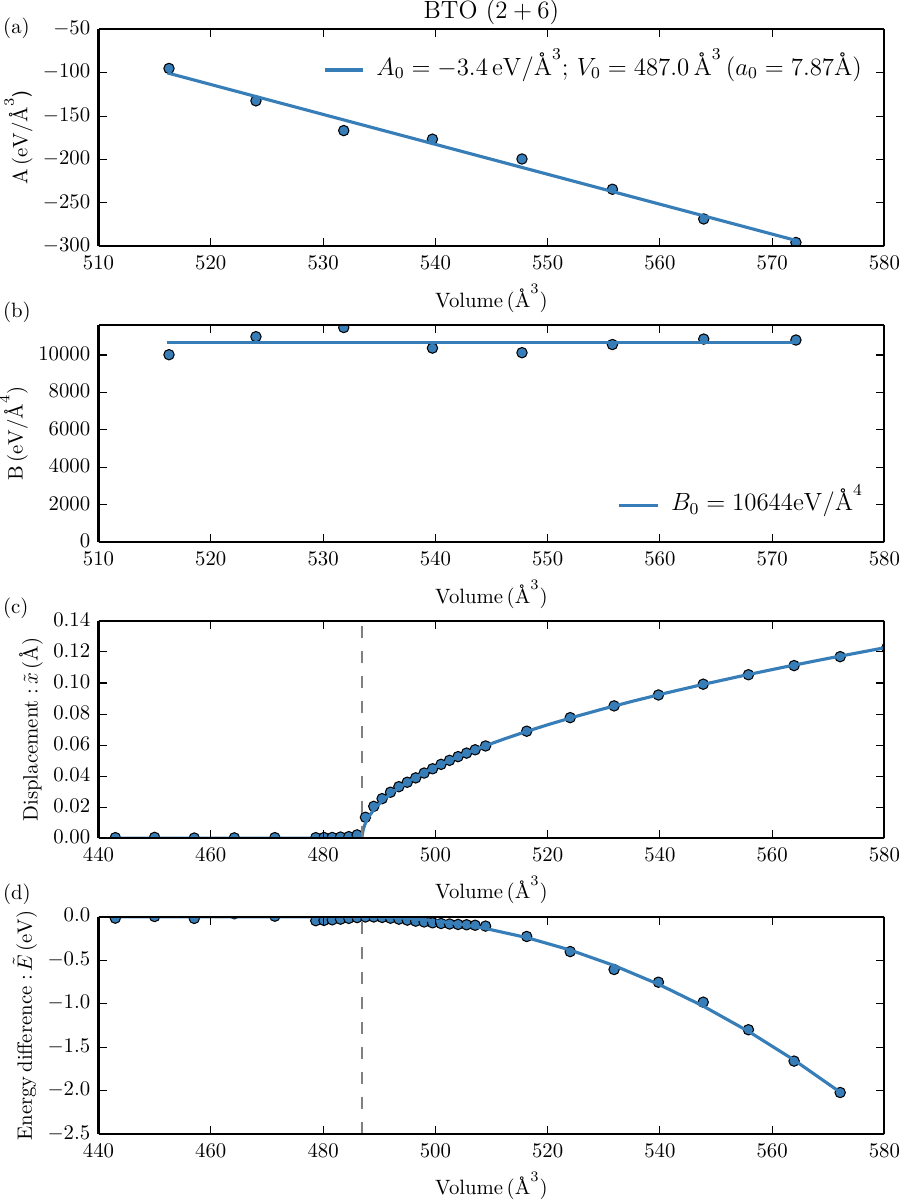}
  \caption{Double-well model for the 2+6 displacement pattern.
  (a) Fit of $A(V)=A_0\cdot (V-V_0)$; (b) fit of $B(V)=B_0$; (c) displacement $\tilde{x}(V)$ and analytical expression using the fit parameters from panels a and b; and, (d) energy difference $\tilde{E}(V)$ of the relaxed displacement patterns w.r.t. the undistorted structure.
  }
  \label{fig:dw2-6}
\end{minipage}
\end{figure}

 From our DFT results of the relaxed displacement patterns as a function of lattice parameter (or, equivalently, volume) we can extract values for $\tilde x(V)$, and the corresponding total energy $\tilde E(V)$, ans thus obtain data points for $A(V)$ and $B(V)$ as a function of the volume from Eq.~\eqref{AB}.
 These are reported in Fig.~\ref{fig:dw4-4}a,b (Fig.~\ref{fig:dw2-6}a,b) for the 4+4 (2+6) displacement pattern. 
 Because of the dependence of the data points, we fit them with the following functional forms: $A(V) = A_0\cdot (V-V_0)$ (with $A_0$ and $V_0$ two constants) and $B(V)$ with a constant $B_0$; the resulting fits are reported in the plots (as well as $a_0=\sqrt[3]{V_0}$, for convenience). 
 In Fig.~\ref{fig:dw4-4}c (Fig.~\ref{fig:dw2-6}c) we then report the data for $\tilde x$ as a function of $\tilde V$ for the 4+4 (2+6) pattern, where the solid line is the analytical expression for $\tilde x(V) = \sqrt{A_0 (V-V_0)/2B_0}$ (for $V > V_0$), obtained using the fitted values for $A_0$, $V_0$ and $B_0$ in panels a and b, that reproduces very well the DFT data points.
 Similarly, in Fig.~\ref{fig:dw4-4}d (Fig.~\ref{fig:dw2-6}d) we plot the energy difference $\tilde E$ of the 4+4 (2+6) pattern with respect to the undistorted structure, and the corresponding analytical curve $\tilde E = -A_0^2(V-V_0)^2/4B_0$ (for $V > V_0$). 

\clearpage
\section{Titanium displacement across the titanate family, and relation to the phonon mode instabilities}
\label{SI:titanates}

\begin{figure}[h]
\centering
\begin{minipage}{.47\textwidth}
  \centering
  \includegraphics[height=0.24\textheight]{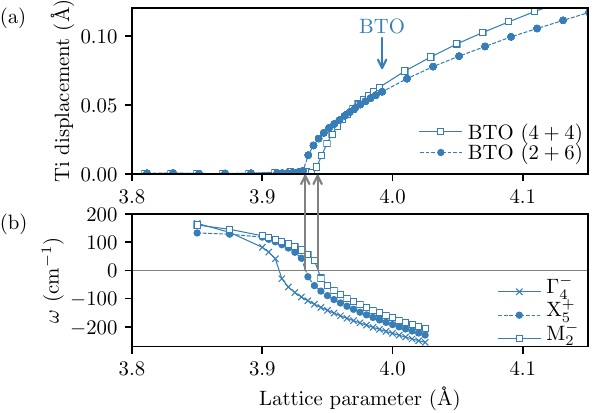}
  \caption{
    Same as Fig.~3~
    in the main text. (a) Magnitude of the displacement of the Ti atoms along one Cartesian coordinate for the 4+4 and 2+6 displacement patterns in  BaTiO$_3$.
(b) Plot of the unstable phonon modes of BaTiO$_3$ with irreducible representation (irrep) $\Gamma_4^-$, X$_5^+$ and M$_2^-$ in the 5-atom undistorted cubic cell as a function of the lattice parameter.}
  \label{fig:SIBTO}
\end{minipage}%
\begin{minipage}{.03\textwidth}
  \centering
  \quad
 \end{minipage}%
\begin{minipage}{.47\textwidth}
\centering
  \includegraphics[height=0.24\textheight]{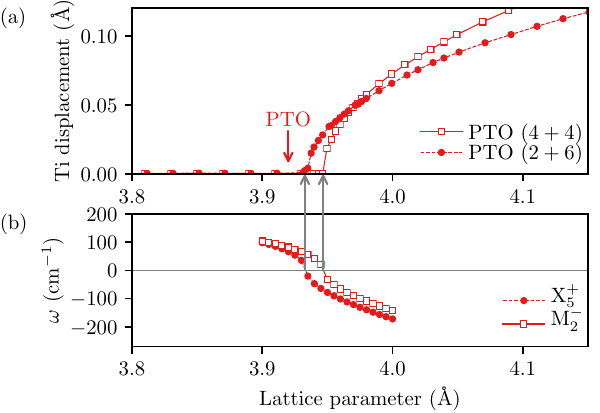}
  \caption{Same analysis for PbTiO$_3$ as in  Fig.~\ref{fig:SIBTO}: (a) Magnitude of the displacement of the Ti atoms along one Cartesian coordinate for the 4+4 and 2+6 displacement patterns in PbTiO$_3$.
(b) Plot of the unstable phonon modes of PbTiO$_3$ with irrep X$_5^+$ and M$_2^-$ in the 5-atom undistorted cubic cell as a function of the lattice parameter. }
  \label{fig:SIPTO}
\end{minipage}
\end{figure}
\begin{figure}[h]
  \centering
  \begin{minipage}{.47\textwidth}
  \centering
  \includegraphics[width=\linewidth]{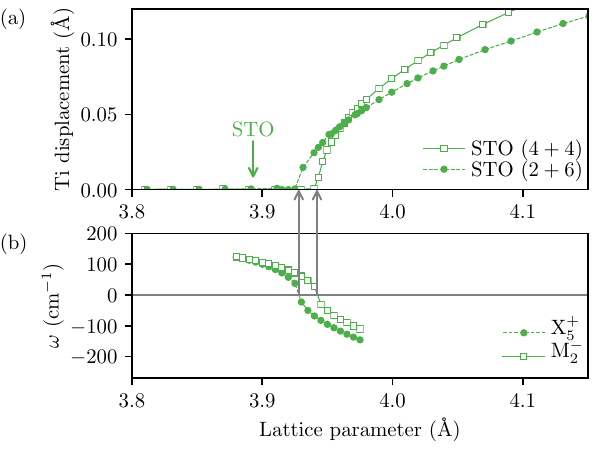}
  \caption{Same analysis for SrTiO$_3$ as in  Fig.~\ref{fig:SIBTO}: (a) Magnitude of the displacement of the Ti atoms along one Cartesian coordinate for the 4+4 and 2+6 displacement patterns in SrTiO$_3$.
(b) Plot of the unstable phonon modes of SrTiO$_3$ with irrep X$_5^+$ and M$_2^-$ in the 5-atom undistorted cubic cell as a function of the lattice parameter.\color{white}{The nonmonotonicity of the Ti displa}}
  \label{fig:SISTO}
  \end{minipage}%
  \begin{minipage}{.03\textwidth}
  \centering
  \quad
 \end{minipage}%
\begin{minipage}{.47\textwidth}
  \centering
  \includegraphics[width=\linewidth]{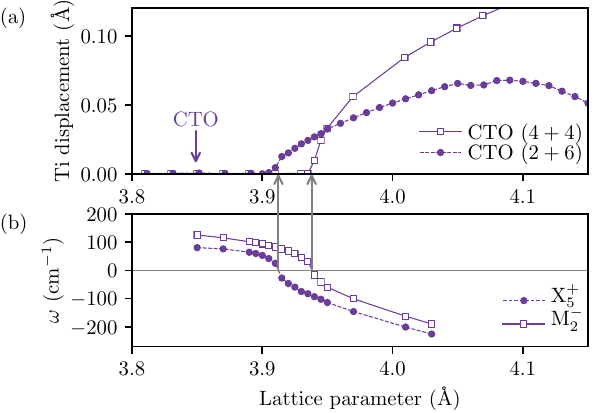}
  \caption{Same analysis for CaTiO$_3$ as in  Fig.~\ref{fig:SIBTO}: (a) Magnitude of the displacement of the Ti atoms along one Cartesian coordinate for the 4+4 and 2+6 displacement patterns in CaTiO$_3$.
(b) Plot of the unstable phonon modes of CaTiO$_3$ with irrep X$_5^+$ and M$_2^-$ in the 5-atom undistorted cubic cell as a function of the lattice parameter. The non-monotonicity of the Ti displacement for the 2+6 ordering is discussed in Fig.~\ref{fig:SICTO26}.}
  \label{fig:SICTO}
\end{minipage}
\end{figure}
We explore the stability of the 4+4 and 2+6 displacement patterns for other perovskite oxides of the titanate family, namely PbTiO$_3$,  SrTiO$_3$, and  CaTiO$_3$.
For the sake of completeness, we include here also Fig.~\ref{fig:SIBTO} with the results for \bto{}, i.e., the same as Fig.~3 
of the main text.
These figures show that the 4+4 (2+6) displacement pattern onset coincides with the lattice parameter at which the  M$_2^-$ (X$_5^+$) irrep  of the 5-atom undistorted cubic cell becomes unstable, indicated by the gray arrows.
For clarity, these figures do not include other zone-boundary phonon modes that are unstable within this range of lattice parameters; however, they are discussed in the material-specific sections below. 
The $\Gamma_4^-$ irrep does not have any cubic subgroups and, therefore, does not correspond to a metastable cubic structure, so we omit it from the other plots.
\subsection{Lead titanate}
\vspace{-10pt}
At the equilibrium lattice constant of cubic PbTiO$_3$ computed with PBEsol (3.93\AA), we find unstable zone-boundary phonons at $q=$ M and R as well as an unstable zone-center mode. These modes, omitted from Fig.~\ref{fig:SIPTO} as they are not involved in the 4+4 or 2+6 displacement patterns, transform with the symmetry of irreps M$^+_2$, R$^-_5$, and $\Gamma^-_4$, respectively. The irreps at R and $\Gamma$ have no cubic subgroups, and the mode that transforms as M$^+_2$ is dominated by oxygen. Using instead a lattice constant of 3.97\AA{} (the experimental cubic lattice parameter for the cubic phase, stable at temperatures higher than 763K~\cite{Jona}), we find additional unstable phonons at $q=$ X and M that transform like  X$^+_5$ and M$^-_2$, respectively (like for \bto{}).
These are the modes that are present in the 2+6 and 4+4 displacement patterns, respectively.
As for \bto{}, the Ti displacement onset for the 2+6 (4+4) displacement patterns correlates with the X$^+_5$ (M$^-_2$) mode becoming unstable, see arrows in Fig.~\ref{fig:SIPTO}.
The stability of the 4+4 and 2+6 displacement patterns at the experimental lattice parameter of cubic PbTiO$_3$ suggests that the Ti atoms will displace along the~\one~directions and is consistent with the non-negligible degree of order-disorder that has been observed experimentally~\cite{Sicron1994,Sicron1995,Sato2005,Fang2015,Yoshiasa2016}.
As we continue to increase the lattice constant we find that additional modes become unstable, in particular the ones transforming like M$^-_5$ (at $\sim$3.975\AA) and like X$^-_5$ (at $\sim$4.00\AA). 
\subsection{Strontium titanate}
\vspace{-10pt}
At the equilibrium lattice constant of cubic SrTiO$_3$ computed with PBEsol (3.89\AA), we find unstable zone-boundary phonons at $q=$ M and R as well as an unstable zone-center mode.
These modes, again  omitted from Fig.~\ref{fig:SISTO} as they are not involved in the 4+4 or 2+6 displacement patterns, transform with the symmetry of irreps M$^+_2$, R$^-_5$, and $\Gamma^-_4$ respectively.
Again we show that the Ti displacement onset for the 2+6 (4+4) displacement patterns correlates with the X$^+_5$ (M$^-_2$) mode becoming unstable, see arrows in Fig.~\ref{fig:SISTO}.
\subsection{Calcium titanate}
\label{sec:CTO-behavior}
\vspace{-10pt}

At the equilibrium lattice constant of cubic CaTiO$_3$ computed with PBEsol (3.85\AA), we find unstable zone-boundary phonons at $q=$ M and R as well as an unstable zone-center mode.
These modes, again  omitted from Fig.~\ref{fig:SICTO} as they are not involved in the 4+4 or 2+6 displacement patterns, transform with the symmetry of irreps M$^+_2$, M$^-_5$ R$^-_5$, and $\Gamma^-_4$ respectively.
The displacement pattern associated with R$_5^-$ has no cubic subgroups, while the unstable M modes do correspond with cubic subgroups. The displacement pattern associated with M$_2^+$ only involves motion of the oxygen atoms; the displacement pattern associated with  M$_5^-$ involves all of the atoms; however, the displacements of the calcium and oxygen are more pronounced than the displacement of the titanium. This implies that, in the paraelectric phase, displacements from the high-symmetry structure would be dominated by the oxygen and calcium with only a small contribution from the titanium. This is not surprising as the tolerance factor of CaTiO$_3$ is well below 1 and has a well-known $Pbnm$ (spacegroup 62) orthorhombic phase that can be constructed through a linear combination of the M$_2^+$ and R$_5^-$ modes.  It is worth noting that, according to the M$_5^-$ mode, the titanium atoms are still restricted by symmetry to displace along the local $\langle$111$\rangle$ directions.
Again we show that the Ti displacement onset for the 2+6 (4+4) displacement patterns correlates with the X$^+_5$ (M$^-_2$) mode becoming unstable, see arrows in Fig.~\ref{fig:SICTO}. However, we see that the onset for the 2+6 pattern  occurs at a smaller lattice parameter than the other titanates. Moreover, we see that as the lattice parameter increases, the titanium displacement reaches a maximum and then decreases, unlike all the other titanium displacement curves of the other titanates.

\begin{figure}[h!]
  \centering
    \includegraphics[width=.5\linewidth]{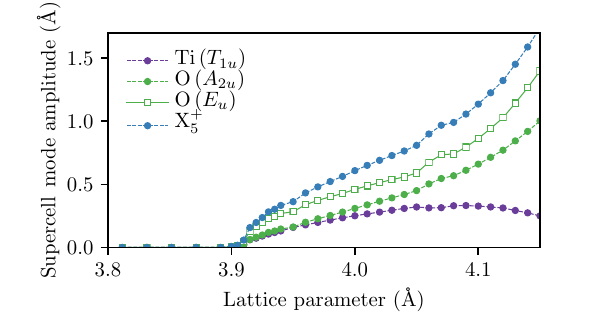}
    \caption{The supercell mode amplitude for the X$^+_5$ irrep and its Ti and O components present in the 2+6 (40-atom) structures.}
 \label{fig:SICTO26}
\end{figure}

In contrast to the behavior of the Ti displacement as a function of the lattice parameter, the X$^+_5$ mode of the 5-atom primitive cell continues to soften monotonically,  while at the same time, the other mode present in the structure (M$^+_5$, which only displaces oxygen atoms), remains stable. Using ISODISTORT to analyze the symmetry of the distortion present in the 2+6 displacement patterns under larger tensile strain, we find that the displacement associated with the X$^+_5$ irrep (a six-dimensional irrep) becomes dominated by oxygen. In Fig.~\ref{fig:SICTO26} we plot the supercell mode amplitude as a function of lattice parameter for the  X$^+_5$ irrep present in the 2+6 (40-atom) structures.
In green are the supercell mode amplitudes associated with the oxygen displacements that transform according to the $A_{2u}$ (solid circles) and $E_u$ (open squares) irreps of the point-group and in purple is the supercell mode amplitude associated with the titanium displacement.
The total supercell X$_5^+$ mode amplitude, in blue, is the norm of the three components, i.e., the square root of the sum of each component squared. 
To convert from supercell mode amplitude $A_{s,\mathrm{Ti}}$ plotted in Fig.~\ref{fig:SICTO26} (as defined within the ISODISTORT~\cite{isodistort,Campbell2006} program) to the titanium displacement $\Delta x_{\mathrm{Ti}}$ of Fig.~\ref{fig:SICTO} (the displacement in \AA~from the cubic high-symmetry position along one coordinate), the following prescription is used:
\begin{align*}
  \Delta x_{\mathrm{Ti}} &= A_{s,\mathrm{Ti}}\cdot a_s\cdot n = A_{s,\mathrm{Ti}}/\sqrt{m},\\
  n &= 1/\sqrt{m\cdot a_s^2},
\end{align*}
where $a_s$ is the supercell lattice parameter (\textit{i.e.,} the lattice parameter of the 2$\times$2$\times$2 supercell), $n$ is a normalization factor, and $m$ is the number of components of the displacement pattern associated with the irrep. 
Here, the number components for  the titanium displacements is 24: the  8 Ti atoms displace along all three components, so we have $m = 8\cdot 3 = 24$. 

  \section{Stability of the 4+4 and 2+6 microscopic templates across a family of 49 perovskites at relaxed volume}
  \quad
  
  \label{SI:pero}

\begin{figure*}[h!]
  \center
    \includegraphics[width=\linewidth]{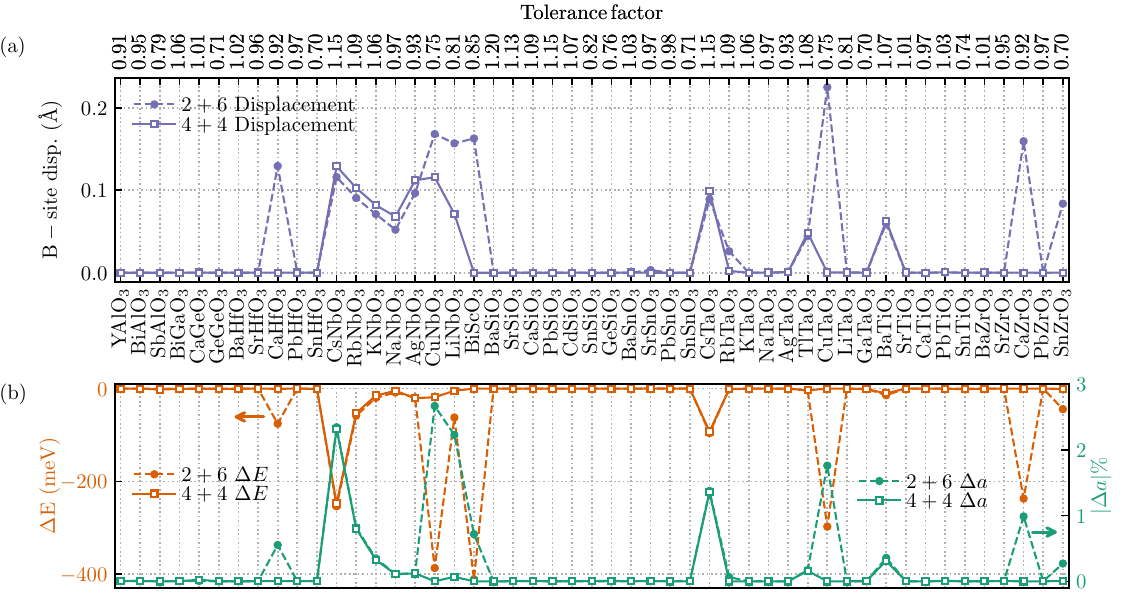}
  
    \caption{
      Occurrence of cubic-symmetry displacement patterns in $2\times 2 \times 2$ supercells for the family of 49 perovskites from Armiento \textit{et al.}~\cite{Armiento2014}. Solid line: 4+4 displacement pattern; dashed line: 2+6 displacement pattern. The lattice parameter is fixed to that of the relaxed 5-atom cubic structure. The tolerance factor is given along the top axis
    (a) Off-site displacement of the B-site cation.
    (b) Energy difference per formula unit $\Delta E$ and relative lattice parameter change $|\Delta a|$, both with respect to the cubic cell without displacements.}
  \label{fig:SIdispE}
\end{figure*}
To investigate the universality of the paraelectric phases in other cubic perovskites, we explore the stability of the 4+4 and 2+6 orderings for the entire set of perovskites in the work of Armiento \textit{et al.}~\cite{Armiento2014}.
The Goldschmidt tolerance factor $t$~\cite{Goldschmidt1926, Jona}, defined as
\begin{align*}
    t = \frac{r_A+r_O}{\sqrt{2}\left(r_B+r_O \right)}, 
\end{align*}
where $r_i$ are the atomic radii of the A-site cation, the B-site cation or the oxygen anion. We calculate the tolerance factor using the ionic radii from Ref.~\onlinecite{Shannon1976}~taken from the database hosted on Ref.~\onlinecite{radii}. 
In cases where the ionic radii for the relevant oxidation state or coordination number was not available, we used the closest one. 

In materials where the 4+4 and 2+6 prototypes are significantly lower in energy, the tolerance factor is $t \gtrsim$ 1.1, and in materials where only the  only the 2+6 prototype significantly lowers the energy, the tolerance factor is $t \lesssim$ 0.9 (Fig.~\ref{fig:SIdispE}).
Specifically, Cs-based compounds have $t> 1.1$;  Rb-based compounds have $t\approx 1.1$;  Ca-based compounds have $t\approx 0.84-0.9$; Li-based compounds have $t\approx 0.8$; Cu-based compounds have $t < 0.8$; and BiScO$_3$ has $t\approx 0.7$.
Note that a number of compounds stabilize the 2+6 and 4+4 prototypes, but the energy gain is quite modest, such as~\bto~or KNbO$_3$. In these cases we do not find a general trend of the tolerance factors. Move over, many of the silicates, that have a tolerance factor less than 0.9 or greater than 1.1, do not stabilize neither the 4+4 nor the 2+6 orderings further supporting the fact that we cannot predict what prototypes will be stable or not from the tolerance factor alone. 

We find that these large energy gains correlate with the change in lattice parameter between the cubic cell without displacements and the one with displacements; however, the energy gain is not due to the change in lattice parameter alone, and a significant energy gain occurs even when the total volume is kept fixed. A better understanding of the behavior of these systems can be obtained by investigating the occurrence of stable displacements as a function of the system volume, see Fig.~\ref{fig:SI-all-disp}.

\section{Stability of 4+4 and 2+6 orderings in the niobates, titanates and zirconates as a function of volume}
\label{SI:NbZr}
\begin{figure}[h!]
  \centering
    \includegraphics[width=\linewidth]{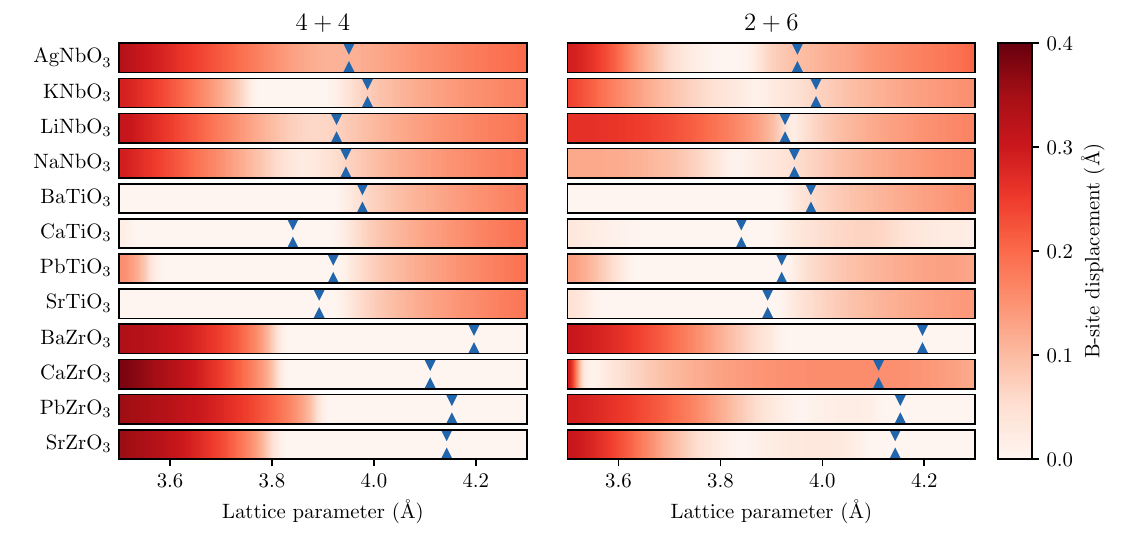}
    \caption{The B-site displacement for the 4+4 and 2+6 displacement patterns for selected members of the niobates, titanates and zirconates as a function of the lattice parameter. The titanates are included here as well for comparison (same results reported earlier in Sec.~\ref{SI:titanates}, here  with a different graphical representation). The equilibrium lattice parameter (calculated with PBEsol) is indicated by the blue markers. \label{fig:SI-all-disp}
    }
\end{figure}

We investigate 12 different perovskites as a function of volume, focusing only on the 4+4 and 2+6 displacement patterns, to demonstrate the generality of the occurrence of local displacements in different perovskite families. The results are shown in Fig.~\ref{fig:SI-all-disp}

The phonon dispersion of cubic 5-atom KNbO$_3$, which also stabilizes the 4+4 and 2+6 displacement patterns, has the same zone-center and zone-boundary instabilities~\cite{Yu1995} as \bto at the equilibrium volume;
however, the Nb off-centering as a function of volume does not match that of the titanates.
In particular, for all niobates that we investigate here, also a volume compression can trigger the occurrence of local displacements, and there is a very narrow lattice parameter range (or even no range at all) in which displacements do not occur. This occurs as additional phonons become unstable under compression.
We also present four zirconates, which have a different B-site displacement vs.{} volume relationship, where instead a compressive strain is needed to stabilize these two patterns, and at relaxed volume no displacements are stabilized (except for CaZrO$_3$ in the 2+6 pattern, where also an increasing volume stabilizes them, including at the relaxed volume).
In all cases, we confirm that the volume dependence and stability of the structural prototypes in a given material cannot be inferred from equilibrium properties alone (like in Fig.~\ref{fig:SIdispE}), but a detailed understanding requires an analysis as a function of the system volume. 

\vspace{1in}

  \section{Titanium-displacement dependence on DFT functional}
  The choice of the DFT functional essentially does not affect the magnitude of Ti displacements, as well as the onset as a function of the lattice parameter, for both the 4+4 and 2+6 displacement patterns; however, it does change the predicted equilibrium lattice constant (indicated by vertical arrows). In the figure we compare the PBEsol~\cite{Perdew2008}, PBE~\cite{Perdew1996,Perdew1997}, and LDA~\cite{KS65} functionals. The PBEsol equilibrium lattice parameter (at $T=0$~K) is the closest to the experimental lattice constant (at finite temperature) of $\sim$ 4.01\AA~\cite{Jona,Bland1959} of cubic BaTiO$_3$ (stable above $\sim$390~K). 
  \quad
  \label{SI:func}
\begin{figure}[h]
  \centering
  \includegraphics[width=.5\linewidth]{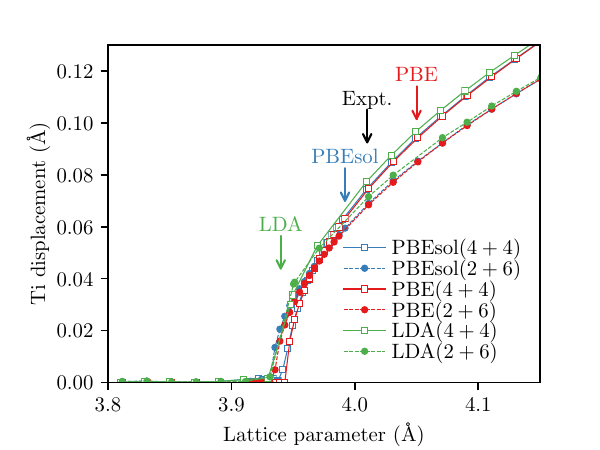}
  
  \caption{ Onset of the cubic-phase instabilities as a function of the lattice parameter for the 4+4 and 2+6 structures of \bto~using the PBEsol, PBE and LDA functionals. The relaxed lattice parameter for each functional in its lowest-energy cubic configuration is indicated by an arrow. }

\end{figure}

\putbib
\end{bibunit}

\begin{thebibliography}{67}
\bibitem [{not()}]{note1}%
  \BibitemOpen
  \href@noop {} {}\bibinfo {note} {We stress here that, as we refer to
  different supercell sizes throughout this paper, we use the term lattice
  parameter to indicate the average lattice parameter of an equivalent 5-atom
  primitive cell: e.g., the actual supercell lattice parameter divided by 4 for
  a 4$\times$4$\times$4 supercell.}\BibitemShut {Stop}%
\end{thebibliography}
\end{document}